
\catcode`@=11
\expandafter\ifx\csname inp@t\endcsname\relax\let\inp@t=\input
\def\input#1 {\expandafter\ifx\csname #1IsLoaded\endcsname\relax
\inp@t#1%
\expandafter\def\csname #1IsLoaded\endcsname{(#1 was previously loaded)}
\else\message{\csname #1IsLoaded\endcsname}\fi}\fi
\catcode`@=12

\font\twelverm=cmr12			\font\twelvei=cmmi12
\font\twelvesy=cmsy10 scaled 1200	\font\twelveex=cmex10 scaled 1200
\font\twelvebf=cmbx12			\font\twelvesl=cmsl12
\font\twelvett=cmtt12			\font\twelveit=cmti12
\font\twelvesc=cmcsc10 scaled 1200	\font\twelvesf=cmss12
                     
\font\twelvemib=cmmib10 scaled 1200
\font\tenmib=cmmib10
\font\eightmib=cmmib10 scaled 800

\skewchar\twelvei='177			\skewchar\twelvesy='60
\skewchar\twelvemib='177

\newfam\mibfam

\def\twelvepoint{\normalbaselineskip=12.4pt plus 0.1pt minus 0.1pt
  \abovedisplayskip 12.4pt plus 3pt minus 9pt
  \belowdisplayskip 12.4pt plus 3pt minus 9pt
  \abovedisplayshortskip 0pt plus 3pt
  \belowdisplayshortskip 7.2pt plus 3pt minus 4pt
  \smallskipamount=3.6pt plus1.2pt minus1.2pt
  \medskipamount=7.2pt plus2.4pt minus2.4pt
  \bigskipamount=14.4pt plus4.8pt minus4.8pt
  \def\rm{\fam0\twelverm}          \def\it{\fam\itfam\twelveit}%
  \def\sl{\fam\slfam\twelvesl}     \def\bf{\fam\bffam\twelvebf}%
  \def\mit{\fam 1}                 \def\cal{\fam 2}%
  \def\sc{\twelvesc}		   \def\tt{\twelvett}%
  \def\sf{\twelvesf}               \def\mib{\fam\mibfam\twelvemib}%
  \textfont0=\twelverm   \scriptfont0=\tenrm   \scriptscriptfont0=\sevenrm
  \textfont1=\twelvei    \scriptfont1=\teni    \scriptscriptfont1=\seveni
  \textfont2=\twelvesy   \scriptfont2=\tensy   \scriptscriptfont2=\sevensy
  \textfont3=\twelveex   \scriptfont3=\twelveex\scriptscriptfont3=\twelveex
  \textfont\itfam=\twelveit
  \textfont\slfam=\twelvesl
  \textfont\bffam=\twelvebf \scriptfont\bffam=\tenbf
                            \scriptscriptfont\bffam=\sevenbf
  \textfont\mibfam=\twelvemib \scriptfont\mibfam=\tenmib
                              \scriptscriptfont\mibfam=\eightmib
  \normalbaselines\rm}


\mathchardef\alpha="710B
\mathchardef\beta="710C
\mathchardef\gamma="710D
\mathchardef\delta="710E
\mathchardef\epsilon="710F
\mathchardef\zeta="7110
\mathchardef\eta="7111
\mathchardef\theta="7112
\mathchardef\iota="7113
\mathchardef\kappa="7114
\mathchardef\lambda="7115
\mathchardef\mu="7116
\mathchardef\nu="7117
\mathchardef\xi="7118
\mathchardef\pi="7119
\mathchardef\rho="711A
\mathchardef\sigma="711B
\mathchardef\tau="711C
\mathchardef\phi="711E
\mathchardef\chi="711F
\mathchardef\psi="7120
\mathchardef\omega="7121
\mathchardef\varepsilon="7122
\mathchardef\vartheta="7123
\mathchardef\varpi="7124
\mathchardef\varrho="7125
\mathchardef\varsigma="7126
\mathchardef\varphi="7127


\def\beginlinemode{\endmode
  \begingroup\parskip=0pt \obeylines\def\\{\par}\def\endmode{\par\endgroup}}
\def\beginparmode{\endmode
  \begingroup \def\endmode{\par\endgroup}}
\let\endmode=\par
{\obeylines\gdef\
{}}
\def\singlespace{\baselineskip=\normalbaselineskip}

\def\oneandahalfspace{\baselineskip=\normalbaselineskip
  \multiply\baselineskip by 3 \divide\baselineskip by 2}
\def\doublespace{\baselineskip=\normalbaselineskip \multiply\baselineskip by 2}

\newcount\firstpageno
\firstpageno=2
\footline={\ifnum\pageno<\firstpageno{\hfil}\else{\hfil\twelverm\folio\hfil}\fi}
\def\toppageno{\global\footline={\hfil}\global\headline
  ={\ifnum\pageno<\firstpageno{\hfil}\else{\hfil\twelverm\folio\hfil}\fi}}
\let\rawfootnote=\footnote		
\def\footnote#1#2{{\rm\singlespace\parindent=0pt\parskip=0pt
  \rawfootnote{#1}{#2\hfill\vrule height 0pt depth 6pt width 0pt}}}
\def\raggedcenter{\leftskip=4em plus 12em \rightskip=\leftskip
  \parindent=0pt \parfillskip=0pt \spaceskip=.3333em \xspaceskip=.5em
  \pretolerance=9999 \tolerance=9999
  \hyphenpenalty=9999 \exhyphenpenalty=9999 }
\def\dateline{\rightline{\ifcase\month\or
  January\or February\or March\or April\or May\or June\or
  July\or August\or September\or October\or November\or December\fi
  \space\number\year}}
\def\received{\vskip 3pt plus 0.2fill
 \centerline{\sl (Received\space\ifcase\month\or
  January\or February\or March\or April\or May\or June\or
  July\or August\or September\or October\or November\or December\fi
  \qquad, \number\year)}}


\hsize=6.5truein
\vsize=8.9truein
\def\nooffset{\global\hoffset=0pt\global\voffset=0pt}

\nooffset
\parskip=\medskipamount
\def\\{\cr}
\twelvepoint		
\doublespace		
\overfullrule=0pt	




\def\title			
  {\null\vskip 3pt plus 0.2fill
   \beginlinemode \doublespace \raggedcenter \bf}

\def\author			
  {\vskip 3pt plus 0.2fill \beginlinemode
   \singlespace \raggedcenter\rm}

\def\affil			
  {\beginlinemode\oneandahalfspace \raggedcenter \sl}

\def\abstract			
  {\vskip 3pt plus 0.3fill \beginparmode
   \oneandahalfspace ABSTRACT: }

\def\endtitlepage		
  {\endpage			
   \body}
\let\endtopmatter=\endtitlepage

\def\body			
  {\beginparmode}		

\def\head#1{			
  \goodbreak\vskip 0.4truein	
  {\immediate\write16{#1}
   \raggedcenter {\sc #1}\par}
   \nobreak\vskip 0truein\nobreak}

\def\itemitemitem{\par\indent\indent \hangindent3\parindent \textindent}
\def\itemitemitemitem{\par\indent\indent\indent \hangindent4\parindent
\textindent}
\def\beginitems{\par\medskip\bgroup
  \def\i##1 {\par\noindent\llap{##1\enspace}\ignorespaces}%
  \def\ii##1 {\item{##1}}%
  \def\iii##1 {\itemitem{##1}}%
  \def\iiii##1 {\itemitemitem{##1}}%
  \def\iiiii##1 {\itemitemitemitem{##1}}
  \leftskip=36pt\parskip=0pt}\def\enditems{\par\egroup}

\def\makefigure#1{\parindent=36pt\item{}Figure #1}

\def\figure#1 (#2) #3\par{\goodbreak\midinsert
\vskip#2
\bgroup\makefigure{#1} #3\par\egroup\endinsert}

\def\beneathrel#1\under#2{\mathrel{\mathop{#2}\limits_{#1}}}

\def\refto#1{$^{#1}$}		

\def\references			
  {\head{References}		
   \beginparmode
   \frenchspacing \parindent=0pt \leftskip=1truecm
   \parskip=8pt plus 3pt \everypar{\hangindent=\parindent}}

\gdef\refis#1{\item{#1.\ }}			

\gdef\journal#1, #2, #3, 1#4#5#6{		
    {\sl #1~}{\bf #2}, #3 (1#4#5#6)}		

\def\refstylenp{		
  \gdef\refto##1{ [##1]}			
  \gdef\refis##1{\item{##1)\ }}			
  \gdef\journal##1, ##2, ##3, ##4 {		
     {\sl ##1~}{\bf ##2~}(##3) ##4 }}

\def\prd{\journal Phys. Rev. D, }

\def\npb{\journal Nucl. Phys., B}

\def\pl{\journal Phys. Lett., }

\def\endreferences{\body}

\def\figurecaptions		
  {\endpage
   \beginparmode
   \head{Figure Captions}
}

\def\endpage			
  {\vfill\eject}

\def\endpaper			
  {\endmode\vfill\supereject}

\def\heading				
  {\vskip 0.5truein plus 0.1truein	
   \beginparmode \def\\{\par} \parskip=0pt \singlespace \raggedcenter}

\def\subheading				
  {\vskip 0.25truein plus 0.1truein	
   \beginlinemode \singlespace \parskip=0pt \def\\{\par}\raggedcenter}

\def\tag#1$${\eqno(#1)$$}

\def\align#1$${\eqalign{#1}$$}

\def\aligntag#1$${\gdef\tag##1\\{&(##1)\cr}\eqalignno{#1\\}$$
  \gdef\tag##1$${\eqno(##1)$$}}

\def\endaligntag{}

\def\overset #1\to#2{{\mathop{#2}\limits^{#1}}}
\def\underset#1\to#2{{\let\next=#1\mathpalette\undersetpalette#2}}
\def\undersetpalette#1#2{\vtop{\baselineskip0pt
\ialign{$\mathsurround=0pt #1\hfil##\hfil$\crcr#2\crcr\next\crcr}}}

\def\ref#1{Ref.~#1}			
\def\Ref#1{Ref.~#1}			
\def\[#1]{[\cite{#1}]}
\def\cite#1{{#1}}
\def\(#1){(\call{#1})}
\def\call#1{{#1}}
\def\taghead#1{}
\def\frac#1#2{{\textstyle {#1 \over #2}}}
\def\half{{\frac 12}}

\def\12{{1\over2}}

\def\sla{\raise.15ex\hbox{$/$}\kern-.57em}
\def\leaderfill{\leaders\hbox to 1em{\hss.\hss}\hfill}
\def\twiddle{\lower.9ex\rlap{$\kern-.1em\scriptstyle\sim$}}
\def\bigtwiddle{\lower1.ex\rlap{$\sim$}}
\def\gtwid{\mathrel{\raise.3ex\hbox{$>$\kern-.75em\lower1ex\hbox{$\sim$}}}}
\def\ltwid{\mathrel{\raise.3ex\hbox{$<$\kern-.75em\lower1ex\hbox{$\sim$}}}}
\def\square{\kern1pt\vbox{\hrule height 1.2pt\hbox{\vrule width 1.2pt\hskip 3pt
   \vbox{\vskip 6pt}\hskip 3pt\vrule width 0.6pt}\hrule height 0.6pt}\kern1pt}
\def\tdot#1{\mathord{\mathop{#1}\limits^{\kern2pt\ldots}}}
\def\happyface{%
$\bigcirc\rlap{\lower0.3ex\hbox{$\kern-0.85em\scriptscriptstyle\smile$}%
\raise0.4ex\hbox{$\kern-0.6em\scriptstyle\cdot\cdot$}}$}
\def\sadface{%
$\bigcirc\rlap{\lower0.25ex\hbox{$\kern-0.85em\scriptscriptstyle\frown$}%
\raise0.43ex\hbox{$\kern-0.6em\scriptstyle\cdot\cdot$}}$}

\def\pmb#1{\setbox0=\hbox{#1}%
  \kern-.025em\copy0\kern-\wd0
  \kern  .05em\copy0\kern-\wd0
  \kern-.025em\raise.0433em\box0 }

\def\tr{{\rm tr}}
\def\Tr{{\rm Tr}}
\def\exp{{\rm exp}}

\def\vt{Department of Physics\\Virginia Polytechnic Institute\\and State %
University\\Blacksburg VA 24060}


\def\m@th{\mathsurround=0pt}
\def\leftrightarrowfill{$\m@th \mathord\leftarrow \mkern-6mu
 \cleaders\hbox{$\mkern-2mu \mathord- \mkern-2mu$}\hfill
 \mkern-6mu \mathord\rightarrow$}
\def\overleftrightarrow#1{\vbox{ialign{##\crcr
	\leftrightarrowfill\crcr\noalign{\kern-1pt\nointerlineskip}
	$\hfil\displaystyle{#1}\hfil$\crcr}}}
\catcode`@=11
\newcount\tagnumber\tagnumber=0

\immediate\newwrite\eqnfile
\newif\if@qnfile\@qnfilefalse
\def\write@qn#1{}
\def\writenew@qn#1{}
\def\w@rnwrite#1{\write@qn{#1}\message{#1}}
\def\@rrwrite#1{\write@qn{#1}\errmessage{#1}}

\def\taghead#1{\gdef\t@ghead{#1}\global\tagnumber=0}
\def\t@ghead{}

\expandafter\def\csname @qnnum-3\endcsname
  {{\t@ghead\advance\tagnumber by -3\relax\number\tagnumber}}
\expandafter\def\csname @qnnum-2\endcsname
  {{\t@ghead\advance\tagnumber by -2\relax\number\tagnumber}}
\expandafter\def\csname @qnnum-1\endcsname
  {{\t@ghead\advance\tagnumber by -1\relax\number\tagnumber}}
\expandafter\def\csname @qnnum0\endcsname
  {\t@ghead\number\tagnumber}
\expandafter\def\csname @qnnum+1\endcsname
  {{\t@ghead\advance\tagnumber by 1\relax\number\tagnumber}}
\expandafter\def\csname @qnnum+2\endcsname
  {{\t@ghead\advance\tagnumber by 2\relax\number\tagnumber}}
\expandafter\def\csname @qnnum+3\endcsname
  {{\t@ghead\advance\tagnumber by 3\relax\number\tagnumber}}

\def\equationfile{%
  \@qnfiletrue\immediate\openout\eqnfile=\jobname.eqn%
  \def\write@qn##1{\if@qnfile\immediate\write\eqnfile{##1}\fi}
  \def\writenew@qn##1{\if@qnfile\immediate\write\eqnfile
    {\noexpand\tag{##1} = (\t@ghead\number\tagnumber)}\fi}
}

\def\callall#1{\xdef#1##1{#1{\noexpand\call{##1}}}}
\def\call#1{\each@rg\callr@nge{#1}}

\def\each@rg#1#2{{\let\thecsname=#1\expandafter\first@rg#2,\end,}}
\def\first@rg#1,{\thecsname{#1}\apply@rg}
\def\apply@rg#1,{\ifx\end#1\let\next=\relax%
\else,\thecsname{#1}\let\next=\apply@rg\fi\next}

\def\callr@nge#1{\calldor@nge#1-\end-}
\def\callr@ngeat#1\end-{#1}
\def\calldor@nge#1-#2-{\ifx\end#2\@qneatspace#1 %
  \else\calll@@p{#1}{#2}\callr@ngeat\fi}
\def\calll@@p#1#2{\ifnum#1>#2{\@rrwrite{Equation range #1-#2\space is bad.}
\errhelp{If you call a series of equations by the notation M-N, then M and
N must be integers, and N must be greater than or equal to M.}}\else%
 {\count0=#1\count1=#2\advance\count1
by1\relax\expandafter\@qncall\the\count0,%
  \loop\advance\count0 by1\relax%
    \ifnum\count0<\count1,\expandafter\@qncall\the\count0,%
  \repeat}\fi}

\def\@qneatspace#1#2 {\@qncall#1#2,}
\def\@qncall#1,{\ifunc@lled{#1}{\def\next{#1}\ifx\next\empty\else
  \w@rnwrite{Equation number \noexpand\(>>#1<<) has not been defined yet.}
  >>#1<<\fi}\else\csname @qnnum#1\endcsname\fi}

\let\eqnono=\eqno
\def\eqno(#1){\tag#1}
\def\tag#1$${\eqnono(\displayt@g#1 )$$}

\def\aligntag#1\endaligntag
  $${\gdef\tag##1\\{&(##1 )\cr}\eqalignno{#1\\}$$
  \gdef\tag##1$${\eqnono(\displayt@g##1 )$$}}

\def\eqalignno#1{\displ@y \tabskip\centering
  \halign to\displaywidth{\hfil$\displaystyle{##}$\tabskip\z@skip
    &$\displaystyle{{}##}$\hfil\tabskip\centering
    &\llap{$\displayt@gpar##$}\tabskip\z@skip\crcr
    #1\crcr}}

\def\displayt@gpar(#1){(\displayt@g#1 )}

\def\displayt@g#1 {\rm\ifunc@lled{#1}\global\advance\tagnumber by1
        {\def\next{#1}\ifx\next\empty\else\expandafter
        \xdef\csname @qnnum#1\endcsname{\t@ghead\number\tagnumber}\fi}%
  \writenew@qn{#1}\t@ghead\number\tagnumber\else
        {\edef\next{\t@ghead\number\tagnumber}%
        \expandafter\ifx\csname @qnnum#1\endcsname\next\else
        \w@rnwrite{Equation \noexpand\tag{#1} is a duplicate number.}\fi}%
  \csname @qnnum#1\endcsname\fi}

\def\ifunc@lled#1{\expandafter\ifx\csname @qnnum#1\endcsname\relax}

\let\@qnend=\end\gdef\end{\if@qnfile
\immediate\write16{Equation numbers written on []\jobname.EQN.}\fi\@qnend}

\catcode`@=12
\refstylenp
\catcode`@=11
\newcount\r@fcount \r@fcount=0
\def\refreset{\global\r@fcount=0}
\newcount\r@fcurr
\immediate\newwrite\reffile
\newif\ifr@ffile\r@ffilefalse
\def\w@rnwrite#1{\ifr@ffile\immediate\write\reffile{#1}\fi\message{#1}}

\def\writer@f#1>>{}
\def\referencefile{
  \r@ffiletrue\immediate\openout\reffile=\jobname.ref%
  \def\writer@f##1>>{\ifr@ffile\immediate\write\reffile%
    {\noexpand\refis{##1} = \csname r@fnum##1\endcsname = %
     \expandafter\expandafter\expandafter\strip@t\expandafter%
     \meaning\csname r@ftext\csname r@fnum##1\endcsname\endcsname}\fi}%
  \def\strip@t##1>>{}}

\def\citeall#1{\xdef#1##1{#1{\noexpand\cite{##1}}}}
\def\cite#1{\each@rg\citer@nge{#1}}	

\def\each@rg#1#2{{\let\thecsname=#1\expandafter\first@rg#2,\end,}}
\def\first@rg#1,{\thecsname{#1}\apply@rg}	
\def\apply@rg#1,{\ifx\end#1\let\next=\relax
\else,\thecsname{#1}\let\next=\apply@rg\fi\next}

\def\citer@nge#1{\citedor@nge#1-\end-}	
\def\citer@ngeat#1\end-{#1}
\def\citedor@nge#1-#2-{\ifx\end#2\r@featspace#1 
  \else\citel@@p{#1}{#2}\citer@ngeat\fi}	
\def\citel@@p#1#2{\ifnum#1>#2{\errmessage{Reference range #1-#2\space is bad.}%
    \errhelp{If you cite a series of references by the notation M-N, then M and
    N must be integers, and N must be greater than or equal to M.}}\else%
 {\count0=#1\count1=#2\advance\count1
by1\relax\expandafter\r@fcite\the\count0,%
  \loop\advance\count0 by1\relax
    \ifnum\count0<\count1,\expandafter\r@fcite\the\count0,%
  \repeat}\fi}

\def\r@featspace#1#2 {\r@fcite#1#2,}	
\def\r@fcite#1,{\ifuncit@d{#1}
    \newr@f{#1}%
    \expandafter\gdef\csname r@ftext\number\r@fcount\endcsname%
                     {\message{Reference #1 to be supplied.}%
                      \writer@f#1>>#1 to be supplied.\par}%
 \fi%
 \csname r@fnum#1\endcsname}
\def\ifuncit@d#1{\expandafter\ifx\csname r@fnum#1\endcsname\relax}%
\def\newr@f#1{\global\advance\r@fcount by1%
    \expandafter\xdef\csname r@fnum#1\endcsname{\number\r@fcount}}

\let\r@fis=\refis			
\def\refis#1#2#3\par{\ifuncit@d{#1}
   \newr@f{#1}%
   \w@rnwrite{Reference #1=\number\r@fcount\space is not cited up to now.}\fi%
  \expandafter\gdef\csname r@ftext\csname r@fnum#1\endcsname\endcsname%
  {\writer@f#1>>#2#3\par}}

\def\ignoreuncited{
   \def\refis##1##2##3\par{\ifuncit@d{##1}%
     \else\expandafter\gdef\csname r@ftext\csname
r@fnum##1\endcsname\endcsname%
     {\writer@f##1>>##2##3\par}\fi}}

\def\r@ferr{\endreferences\errmessage{I was expecting to see
\noexpand\endreferences before now;  I have inserted it here.}}
\let\r@ferences=\references
\def\references{\r@ferences\def\endmode{\r@ferr\par\endgroup}}

\let\endr@ferences=\endreferences
\def\endreferences{\r@fcurr=0
  {\loop\ifnum\r@fcurr<\r@fcount
    \advance\r@fcurr by 1\relax\expandafter\r@fis\expandafter{\number\r@fcurr}%
    \csname r@ftext\number\r@fcurr\endcsname%
  \repeat}\gdef\r@ferr{}\global\r@fcount=0\endr@ferences}


\let\r@fend=\endpaper\gdef\endpaper{\ifr@ffile
\immediate\write16{Cross References written on []\jobname.REF.}\fi\r@fend}

\catcode`@=12

\citeall\refto		
\citeall\ref		%
\citeall\Ref		%
\nooffset
\title{Regularization Dependence of Quadratic Divergence Cancellations}
\vskip 0.2in
\author{Gary Kleppe}
\vskip 0.2in
\affil\vt
\abstract{Certain results related to the cancellation of quadratic divergences,
which had been obtained using dimensional reduction, are reconsidered using a
nonlocal regulator. The results obtained are shown to depend strongly on the
regulator. Specifically, it is shown that a certain result of Al-sarhi, Jack,
and Jones no longer holds, even if a nontrivial measure factor is used; also
that there are no values of the top and Higgs mass for which the one-loop
quadratic divergence in the standard model cancels independently of the
renormalization scale, whether or not strong interaction effects are ignored.}

\endtopmatter
\nooffset
\head{1. Introduction}

Over the last ten years, there has been some interest\refto{QD,JJ1,JJ2} in
studying the quadratic
divergences of gauge theories, particularly those of the standard model. The
smallness of the Higgs mass is looked at as unnatural in the presence of the
quadratic divergences. From the point of view of standard renormalization
theory it is hard to see what the difficulty is. From this perspective, the
quadratic divergences only exist to be subtracted away. There is not even any
good definition of ``small'', because the cutoff scale is taken to infinity.

One may instead take the viewpoint of Wilson\refto{Wilson}, in which the cutoff
is thought of as some large but finite scale. Field theories such as the
standard model are thought of as effective theories which are only valid
at scales below the cutoff. Above the cutoff scale some unknown new physics
takes over. Thus although this new physics will determine the values of
renormalized parameters, we expect that for quadratically divergent parameters
the values should be of the order of the cutoff scale squared times the
appropriate coupling constant(s); values which are
much smaller are unnatural, as they depend on precise fine-tuning at the new
physics scale. In the case of
the Higgs mass, this would require the new physics scale to be uncomfortably
small, around the 1 TeV range.

One way to avoid this problem is through supersymmetry, which generally causes
quadratic divergences to be cancelled. However, no supersymmetric partners for
known particles have ever been detected experimentally. Therefore it is natural
to look for other theories in which the quadratic divergences cancel.

A curious result concerning quadratic divergences was found by Al-sarhi, Jack,
and Jones\refto{JJ1,JJ2}. They considered the quadratic divergences of a theory
of scalars and fermions using dimensional reduction\refto{DR1} and minimal
subtraction. They found that by demanding that the
quadratic divergences cancel at one loop, and that this cancellation be
unaffected by the action of the renormalization group, cancellation
automatically occured at two loops; and by demanding the two loop
cancellation be invariant under the renormalization group, cancellation at
three
loops was automatic.

It was found that similar results did not occur at the next loop
order\refto{JJ1}, or for gauge theories\refto{JJ2}. However, the gauge theory
result could be fixed if the peculiarities of the dimensional reduction
technique were accounted for. Furthermore, it is known that dimensional
reduction fails at four loops\refto{DR2}, and the renormalization group beta
functions at this order depend on a choice of renormalization prescription,
so it is possible that these failures are due to the choice of regularization
and renormalization scheme. These authors\refto{JJ2} also consider the case of
the standard model\refto{SM}
using the same techniques. They find that there is no
simultaneous soultion of the one and two loop constraints, however the
quadratic divergences at one loop cancel independently of the renormalization
group scale for $m_t=115$ GeV and $m_H=180$ GeV, but only if the strong
interactions' contribution to the beta functions are ignored.

There is another good reason to be suspicious of dimensional
reduction. In this scheme, as in dimensional regularization, one does
not see quadratic divergences directly. They are only inferred by looking at
poles in lower dimensions. At one loop this cannot be argued with, but at two
or more loops it is hard to justify why lower dimensional poles are related to
quadratic divergences. Thus it is worth reexamining these results using a
different regulator, to try to determine just how much they depend on what
techniques are used.

In this paper the question of quadratic divergence cancellation is reexamined
using the technique of nonlocal regularization\refto{NLGT,NLYM} (a version of
which was used at one loop in ref. \cite{JJ2}). In section 2 the theory of
scalars and fermions is considered, and it is shown that the result of
Al-sarhi,
Jack and Jones does not hold with this technique, even if a measure factor is
added to the interactions. In section 3 the standard model is considered, and
it is shown that with the minimal measure factor, there are no real values for
$m_t$ and $m_H$ which make the quadratic divergence cancel independently of the
scale, with or without strong interactions. A nonminimal measure could alter
this result, however.

\head{2. Scalar-Fermi Theory}

\def\psib{\overline\psi}
\def\sigmab{\overline\sigma}
Consider the renormalizable theory of scalars and fermions whose Lagrangian is
$$\eqalign{{\cal L}&=\half\phi_a(\partial^2-m_a^2)\phi_a-\frac16
g_{abc}\phi_a\phi_b
\phi_c-\frac1{24}\lambda_{abcd}\phi_a\phi_b\phi_c\phi_d\cr
&\qquad-i\psib\sigmab\cdot\partial\psi-\half\Bigl[\psi\left(M+Y_a\phi_a\right)
\psi+\hbox{ c.c. }\Bigr]}\eqno(Lsf)$$
where $\psi$ is a two component spinor field, with arbitrary ``flavor'' indices
which we will not show, and $g$ and $\lambda$ are totally symmetric. The metric
used is (--1,1,1,1). Our convention is that repeated scalar indicies are summed
over, even if repeated more than once as in $m_a^2\phi_a^2$.

Nonlocal regularization is described in detail in ref. \cite{NLYM}; detailed
calculational techniques may be found in ref. \cite{Twoloop}.
For our purposes, the technique may be summarized in two steps. The
first step is to expand all propagators in the Schwinger parametrization, e.g.
$${1\over p^2+m^2}=\int_0^\infty {d\tau\over\Lambda^2} \exp\left(-{\tau
(p^2+m^2
)\over\Lambda^2}\right)\eqno()$$
and to delete the regions of Schwinger parameter space for which all parameters
around a closed loop in some diagram are simultaneously less than 1.
The second step is to add a measure factor, a set of
extra interactions which assure that
the path integral measure is invariant under nonlocally distorted versions of
any symmetries which were present in the original theory. In this first simple
case we need
not be concerned with this, since there are no symmetries in this theory which
need to be preserved.

The quadratic divergences in \(Lsf) of course occur in the scalar two-point
function. At one loop there are two contributions to this, shown in figures
1a and 1b.\footnote*{Solid lines denote scalars, dashed lines denote fermions.}
The graph of figure 1a gives a contribution
$$-{\lambda_{abcc}\over 2}\int{d^4 q\over(2\pi)^4}\int_1^\infty{d\tau\over
\Lambda^2}\exp\left[-{\tau\over\Lambda^2}(q^2+m_c^2)\right]$$
$$=-{i\lambda_{abcc}\Lambda^2\over2(4\pi)^2}\int_1^\infty{d\tau\over\tau^2}
\exp\left[-{\tau m_c^2\over\Lambda^2}\right]\eqno(fig1a)$$
\def\lnl{{\rm ln}{\Lambda^2\over\mu^2}}
\def\pf{\hbox{ + finite}}
$$=-{i\lambda_{abcc}\over2(4\pi)^2}(-\Lambda^2+m_c^2\ \lnl)\pf$$
where $\mu$ is an arbitrary scale. The graph of figure 1b gives a contribution
\def\Tr{{\rm Tr}}\def\tr{{\rm tr}}\def\cc{\hbox{ + c.c.}}
\def\loopint#1{\int{d^4 #1\over(2\pi)^4}}
\def\ipi{{i\over(4\pi)^2}}
$$\eqalignno{-&\half \Tr\Biggl\{
Y_a\loopint q\int_{R_1^2}{d\tau_1
d\tau_2\over\Lambda^4}\ \tr\left[(p+q)\cdot\sigmab\ q\cdot\sigma\right]\cr
&\qquad\qquad\times\exp\left[-{1\over\Lambda^2}\left(\tau_1 q^2+\tau_2(p+q)^2+
\tau_{12}M^2\right)\right]\ Y^*_b\cc\Biggr\}&(fig1b)\cr
&=\ipi \Tr\Biggl\{
Y_a\int_{R_1^2}{d\tau_1 d\tau_2\over\tau_{12}^2}
\left({2\Lambda^2\over\tau_{12}}-{\tau_1\tau_2\over\tau_{12}^2}\right)
\exp\left[-{1\over\Lambda^2}\left({\tau_1\tau_2\over\tau_{12}^2} p^2+\tau_{12}
M^2\right)\right]Y^*_b\cc\Biggr\}}$$
where we have used the notations
$$\int_{R_1^n}\equiv\int_0^\infty \ldots\int_0^\infty -\int_0^1\ldots\int_0^1
\eqno()$$
$$\tau_{i_1\ldots i_n}\equiv\tau_{i_1}+\ldots+\tau_{i_n}\eqno()$$
$$\tr\equiv\hbox{ Trace over spinor indices}\eqno()$$
$$\Tr\equiv\hbox{ Trace over other (``flavor'') fermion indices}\eqno()$$
Expanding \(fig1b) we get
$$\half\ipi \Tr\Biggl\{
Y_a\left(3\Lambda^2-p^2\lnl-4M^2\lnl\right)Y^*_b\cc\Biggr\}\pf
\eqno(fig1bresult)$$
Combining \(fig1a) with \(fig1bresult) we find that the one loop quadratic
divergence cancels\footnote*{The scalar one-point function also diverges
quadratically. Cancellation of this would give another condition, this one
involving $g$ and $M$ in addition to $\lambda$ and $Y$. For our purposes it
is not necessary to consider this additional complication.} if
$$\lambda_{abcc}=3\Tr[Y^*_aY_b+Y^*_bY_a]\eqno(oneloopresult)$$
Note that in this and in the following, we neglect many graphs which have no
quadratic divergences. In a similar manner to the above, we obtain the scalar
field strength renormalization
$$\delta Z^\phi_{ab}=-{1\over2(4\pi)^2}\Tr[Y_aY^*_b+Y_a^*Y_b]\
\lnl\eqno(sfsr)$$
where we use a minimal subtraction scheme at scale $\mu$.

Likewise, from the graph of figure 2 we obtain the fermion field
strength renormalization
$$\delta Z^\psi=-\half{\Tr[Y_aY^*_a]\over(4\pi)^2}\ \lnl,\eqno(fsfr)$$
from figure 3 the Yukawa coupling renormalization
$$\delta Y_a={Y_bY_a^*Y_b\over(4\pi)^2}\ \lnl,\eqno(yr)$$
and from figures 4a and 4b the scalar quartic coupling renormalization
$$\eqalign{\delta\lambda_{abcd}&={1\over(4\pi)^2}\Biggl[\half(\lambda_{abef}
\lambda_{cdef}+\lambda_{acef}\lambda_{bdef}+\lambda_{adef}\lambda_{bcef})\cr
&\qquad-2\Tr[Y_a^*Y_bY_c^*
Y_d+Y_a^*Y_cY_b^*Y_d+Y_a^*Y_bY_d^*Y_c\cc]\Biggr]\lnl}\eqno(lamdar)$$
There are of course mass and cubic scalar potential counterterms, but these
will be unnecessary for our purposes. From the above we may calculate the beta
functions for $\lambda$ and $Y$, which agree with the standard results (see
e.g. ref. \cite{JJ2}). Then demanding that ${\partial\over\partial\mu}$ on
\(oneloopresult) gives zero, we get
$$\eqalign{&-\lambda_{abef}\lambda_{ccef}-2\lambda_{acef}\lambda_{bcef}+\Biggl(
16\Tr[Y^*_aY_cY^*_bY_c]+14\Tr[Y^*_aY_bY^*_cY_c]-\half\lambda_{bcdd}\Tr[Y_a^*Y_c]
\cr
&\qquad+3\Tr[Y_aY^*_c](\Tr[Y_bY^*_c]+\Tr[Y_b^*Y_c])-\half\lambda_{acdd}\Tr[
Y_b^*Y_c]\cc\Biggr)-2\lambda_{abcd}\Tr[Y_cY^*_d]=0}\eqno(onelooptwo)$$
Now let us calculate the two loop results for comparison. At two loops there
will be both $\Lambda^2\lnl$ and $\Lambda^2$ divergences. Since these
divergences are always independent of the momenta, we may set the external
momenta to zero to simplify the computations. Let us first consider
the most divergent terms, the $\Lambda^2\lnl$. Only graphs which are one loop
graphs with some internal line corrected by either another one loop graph or a
renormalization will contribute to the leading divergence. The first such graph
is shown in figure 5a. It evaluates to
\def\ipii{{i\over(4\pi)^4}}
\def\tlint{\int{d^4p\ d^4q\over(2\pi)^8}}
$$-{i\lambda_{abcd}\lambda_{cdee}\over 4}\tlint\int_{R_1^2}{d\tau_1d\tau_2
\over\Lambda^4}\int_1^\infty{d\tau_3\over
\Lambda^2}\exp\left[-{1\over\Lambda^2}\left(\tau_{12}p^2+\tau_3q^2+\tau_1m_c^2
+\tau_2m_d^2+\tau_3m_e^2\right)\right]$$
\def\lo{+O(\Lambda^2)}
$$=\ipii{\lambda_{abcd}\lambda_{cdee}\Lambda^2\over4}\lnl\lo\eqno(fig5a)$$
The graph of figure 5b contributes
$$\eqalign{&\frac i2 \lambda_{abcd}\Tr[Y^*_cY_d]\tlint \int{d\tau_1\ldots d
\tau_4\over\Lambda^8} \tr[q\cdot\sigmab(p+q)\cdot\sigma]\cr
&\qquad\times\exp\left[-{1\over\Lambda^2}\left(\tau_{12}p^2+\tau_3(p+q)^2+
\tau_4q^2+\tau_1m_c^2+\tau_2m_d^2\right)\right]}\eqno()$$
where there are two regions of parameter integrations:

\line{$1)\hfill(\tau_1,\tau_2)=R_1^2,\qquad (\tau_3,\tau_4)=R_1^2\hfill$}

and

\line{$2)\hfill\tau_1=(0,1),\qquad\tau_2=(0,1),\qquad\tau_3=(1,\infty),\qquad
\tau_4=(1,\infty)\hfill$}
The second region does not give any divergence at this order; the first does
give an amount
$$\ipii\cdot\frac32\lambda_{abcd}\Tr[Y^*_cY_d]\Lambda^2\lnl\eqno(fig5b)$$
The graph of figure 5c appears to also give a $\Lambda^2\lnl$ contribution, but
is in fact only quadratically divergent. All that remains are the graphs of
figure 6 a-d, which are one loop graphs with renormalization insertions. Figure
6a comes from the quadratic vertex correction. It evaluates to
$$\ipii \left\lbrace-\frac14(\lambda_{abcd}\lambda_{cdee}+2\lambda_{acde}
\lambda_{bcde})+\left(2\Tr[Y_a^*Y_bY_c^*Y_c]+\Tr[Y_a^*Y_cY_b^*Y_c]\cc\right)
\right\rbrace\Lambda^2\lnl\eqno(fig6a)$$
Figure 6b, from the Yukawa renormalization, gives
$$\ipii\cdot 3\ \Tr[Y_aY_c^*Y_bY_c^*\cc]\ \Lambda^2\lnl\eqno(fig6b)$$
The scalar field strength renormalization insertion graph in figure 6c gives
$$-\ipii\cdot\half\lambda_{abcd}\ \Tr[Y_cY_d^*]\ \Lambda^2\lnl\eqno(fig6c)$$
Finally, the fermion field strength renormalization insertion graph in figure
6d gives
$$\ipii\cdot\frac32\Tr[Y_a^*Y_bY_c^*Y_c\cc]\ \Lambda^2\lnl\eqno(fig6d)$$
Putting it all together gives the total
$$\ipii\left(-\half\lambda_{acde}\lambda_{bcde}+\lambda_{abcd}\Tr[Y_c^*Y_d]
+\left\lbrace\frac72\Tr[Y_aY_b^*Y_cY_c^*]+4\Tr[Y_aY_c^*Y_bY^*_c]\cc\right\rbrace
\right)\Lambda^2\lnl\eqno(twoloopresult)$$
Comparing \(twoloopresult) with \(oneloopresult) and \(onelooptwo), we see that
the results do not agree as they did in the dimensional reduction case. Thus
the result in question is seen to fail when nonlocal
regularization is used.

The above discussion assumed that no measure factor was used. As stated
previously, for the scalar-fermi theory considered here there is no need for a
measure factor, but on the other hand neither is there any reason not to
include one. In particular, note that if we choose parameters in \(Lsf) such
that the theory has supersymmetry, it would\refto{SR} be necessary to add a
measure factor to preserve the supersymmetry in loop amplitudes\footnote*{If
supersymmetry auxiliary fields are used, as in the superfield formalism, then
no measure factor is required as in this case supersymmetry is linearly
realized. When such a theory is nonlocally regulated, the auxiliary fields
no longer enter the action purely quadratically. So if one integrates out the
auxiliary fields after nonlocalization, extra interactions are generated which
are the same as the measure factor which would be needed if one just
nonlocalized the theory without the auxiliary fields.}. So let us consider
adding a minimal measure term
$$S_m={1\over(4\pi)^2}\alpha_{ab}\Lambda^2\int d^4x\ \phi_a(x)\ \phi_b(x)
\eqno(sm)$$
where we shall take
$$\alpha_{ab}=A\ \Tr[Y_aY_b^*+Y_a^*Y_b]+B\lambda_{abcc}\eqno()$$
The one loop results \(oneloopresult) and \(onelooptwo) become
$$(B-1)\lambda_{abcc}+(A+3)\ \Tr[Y^*_aY_b+Y_aY^*_b]=0\eqno(oneloopagain)$$
and (after using \(oneloopagain) to eliminate some terms)
$$\eqalign{&(B-1)\left(\lambda_{abef}\lambda_{ccef}+2\lambda_{acef}
\lambda_{bcef}\right)+(-8B-2A+14)\Tr[Y_aY_b^*Y_cY_c^*+Y_a^*Y_bY_c^*Y_c]\cr
&\qquad+(-4B+4A+8)\Tr[Y_a^*Y_cY_b^*Y_c+Y_aY_c^*Y_bY_c^*]
-{(B-1)^2\over A+3}\lambda_{abcd}\lambda_{cdee}}\eqno(onelooptwoagain)$$
The presence of the measure term creates an extra graph at two loops, the one
shown in figure 7. This graph gives an extra contribution to the two loop
result
of
$$\ipii\left[{A(B-1)\over A+3}-B\right]\lambda_{abcd}\lambda_{cdee}\Lambda^2
\lnl\eqno(fig7)$$
giving a new two loop total of
$$\eqalign{&\ipii\Biggl(\biggl[{(A-\half)(B-1)\over A+3}-B\biggr]\lambda_{abcd}
\lambda_{cdee}+\frac72\Tr[Y_a^*Y_bY_c^*Y_c+Y_aY_b^*Y_cY_c^*]\cr
&\qquad+4\Tr[Y_aY_c^*Y_bY_c^*+Y_a^*Y_cY_b^*Y_c]\Biggr)\Lambda^2\lnl}
\eqno(twoloopagain)$$
where we have again used \(oneloopagain). We see that there are no values for
$A$ and $B$ which will make \(twoloopagain) equivalent to \(onelooptwoagain).

There are of course other measure factors that may be added. However, it turns
out that any other scalar measure interaction which does not spoil the
renormalizability of the theory will affect neither \(oneloopagain) or
\(twoloopagain). This is because the measure factor interaction is required
to be analytic in $p^2$. Integrals of the form $\int d^4q f(q)$ may only be
log divergent if $f(q)$ goes like $q^4$ for large $q$. One could consider
adding
additional measure terms involving fermions instead of scalars. Of course such
terms would not affect \(oneloopresult). They could only affect
\(twoloopresult)
if they were of order $\Lambda^2$ or higher. But such terms would also ruin the
power counting behavior of the theory.

So much for the leading order $\Lambda^2\lnl$
divergence. The next order $\Lambda^2$ divergence will be totally dependent
on $\phi^3$ and $\phi^4$ terms in the measure, so it is almost certain that the
result in question could be made to work for these terms with the right choice
of measure. Not only that, but one could take \(oneloopagain) to be true only
to first order, and add higher order terms to this condition which would then
change the condition at second order. None of this, however, will affect the
leading order $\Lambda^2\lnl$ two loop divergences, which clearly do not
satisfy the conjecture of Al-sarhi, Jack, and Jones when nonlocal
regularization
is used.

\head{3. The Standard Model}

In this section we follow the conventions used in ref. \cite{JJ2}. In
particular,
we work in the unbroken phase of the standard model, in which the quadratic
divergences occur in the Higgs self-energy, and neglect all Yukawa couplings
except that of the top quark, which is denoted by $h$. The $SU(2)$ and $U(1)$
couplings are given by $g$ and $g'/2$ respectively. We will use the minimal
measure factor, following the construction outlined in ref. \cite{NLYM}; thus
the measure
factor will not affect the quadratic divergences at one loop. The calculational
procedure then follows that shown in the last section, so we will just give the
results:

The graph of figure 8 gives $-\ipi\cdot\lambda\Lambda^2$.

The graph of figure 9 gives $-\ipi\cdot\frac38 g'{}^2\Lambda^2$.

The graph of figure 10 gives $-\ipi\cdot\frac98 g^2\Lambda^2$.

The graph of figure 11 gives $\ipi\cdot 3h^2\Lambda^2$.

The graph of figure 12 gives $\ipi\cdot g'{}^2\Lambda^2$.

The graph of figure 13 gives $\ipi\cdot 2g^2\Lambda^2$. This graph and the
preceeding one would have been zero in dimensional regularization (or
reduction).

Thus the quadratic divergences will cancel for
$$-\lambda+\frac58g'{}^2+\frac78g^2+3h^2=0\eqno(smonel)$$
Using the standard model beta functions\refto{JJ2}, \(smonel) is independent
of $\mu$ only if
$$-12\lambda^2+(3g'{}^2+9g^2)\lambda+\frac{187}{24}(g'{}^4+g^4)-\frac32g'{}^2g^2
-12\lambda h^2+39h^4-\frac{17}2g'{}^2h^2-\frac{27}2g^2h^2-48g_3^2h^2
\eqno(smonetwo)$$
where $g_3$ is the QCD coupling constant.
Using \(smonel) to eliminate $\lambda$ in \(smonetwo), we get
$$\frac{238}{48}g'{}^4-\frac{51}8g'{}^2g^2-\frac{437}{48}g^4-105h^4-52h^2g'{}^2-
60g^2h^2-48g_3^2h^2=0\eqno(oops)$$
Then using the relations
$$g^2=4m_W^2/v^2,\qquad\qquad g'{}^2=4(m_Z^2-m_W^2)/v^2,\eqno()$$
and the values\refto{slac}
$$m_W=80.6\ {\rm GeV}\qquad\qquad m_Z=91.16\ {\rm GeV},\eqno()$$
we see that \(oops) has no solution for $h^2>0$, regardless of the
value of $g_3$. Thus there are no real values for $m_t$ and $m_h$ which make
the quadratic divergences cancel using this method. Of course, it may be
possible to add a non-minimal measure factor, as in the last section, if one
may do so while maintaining gauge
invariance. If this can be done, then it is likely that \(oops) can be altered
to get whatever answer one would like for the masses.

\head{4. Conclusions}

The purpose of this paper has been to determine how certain results related to
the cancellation of quadratic divergences depend on how a theory is regulated.
We have studied several results of Al-sarhi, Jack, and Jones from dimensional
reduction, and shown that these results are not valid when one instead uses the
technique of nonlocal regularization. With hindsight, we may remark that this
regularization dependence should not have been unexpected. It is true that we
should obtain the same answer independently of a regularization and
renormalization program when we calculate any meaningful physical quantity.
However, the quadratic divergences do not belong to the category of meaningful
physical quantities, so there is no reason not to expect vastly different
answers for quadratic divergences using different methods.\footnote*{If
cancellation of quadratic divergences depends on a regularization scheme, then
why is such cancellation automatic in supersymmetric theories, independently of
any choice of scheme? The answer is that such cancellation requires the use of
a
regularization which respects supersymmetry. This reflects the very reasonable
assumption that if the low energy theory is supersymmetric, then the new
physics
at higher energies will be so as well.}

If quadratic divergences depend strongly on a regularization procedure, then is
there a ``correct'' procedure which gives a meaningfully correct answer, e.g.
which correctly tells us when we need to rely on new physics to keep a
parameter small? The answer seems to be that we should regulate in a way
which approximates physics at scales just below the ``new physics'' cutoff
scale. Unfortunately we do not know what this physics is like. Most would
agree that this physics is probably not described by a dimensionally reduced
theory\footnote{**}{Lower dimensional theories such as ``matrix models'' have
been studied by researchers interested in fundamental physics, however these
theories at most represent toy models and are not by themselves intended to
describe our universe.}. It is more reasonable that a
nonlocal field theory might be appropriate, but the construction used in this
paper is only one of a probable large number of types of nonlocal theories,
and even within this construction there is still ambiguity in the measure.
So at this point there seems to be little reason to trust the results of either
the dimensional reduction method or the nonlocal method.

\head{Acknowledgements}

I thank the authors of ref. \cite{JJ2} for bringing their work to my attention,
and L. N. Chang, E. J. Piard, P. Ramond, P. Sikivie, C. B. Thorn, and R. P.
Woodard for useful information and discussions. This work has been supported
by the United States Department of Energy under contract no.
DOE-AS05-80ER-10713.

\references

\refis{slac} J. J. Hern\'andez {\it et al}, \pl B239, 1990, 1.

\refis{QD}M. Veltman, {\sl Acta Phys. Pol. B}{\bf 12} (1981) 437; T. Inami,
H. Nishino, and S. Watamura, \pl B117, 1982, 197; N. G. Deshpande, R. J.
Johnson, and E. Ma, \pl B130, 1983, 61; \prd 29, 1984, 2851; L. Castellani
and P. van Nieuwenhuizen, \npb 213, 1983, 305; J. Kubo, K. Sibold, and
W. Zimmerman, \pl B220, 1989, 191; Y. Nambu, Chicago preprints EFI 88-62,
88-39, and 90-46; B. Gonzales, M. Ruiz-Altaba, and M. Vargas, preprint CERN
TH. 5558/89; C. Wetterich. {\sl Z. Phys. C}{\bf 48} (1990) 693; M.
Capdequi Peryranere, J. C. Montero, and G. Moultaka, \pl B260, 1991, 138;
M. Vargas and J. L. Lucio, preprint CINVESTAV/FIS-2191.

\refis{SM}The standard model has also been considered in Gonzales et al.,
and Capdequi Peryranere et al., {\sl op cit.}

\refis{NLGT} D. Evens, J. Moffat, G. Kleppe and R. P. Woodard,
\prd 43, 1991, 499.

\refis{NLYM} G. Kleppe and R. P. Woodard, ``Nonlocal Yang-Mills'', preprint
UF-IFT-90-22, Sept. 1990, revised Nov. 1991, to appear in {\sl Nucl. Phys. B}.

\refis{SR} G. Kleppe and R. P. Woodard, \pl B253, 1991, 331.

\refis{DR1}W. Siegel, \pl 84B, 1979, 193.

\refis{DR2}W. Siegel, \pl 94B, 1980, 37.

\refis{Wilson}K. G. Wilson, {\sl Phys. Rev. B}{\bf 4} (1971) 3174, 3184;
K. G. Wilson and J. G. Kogut, {\sl Phys. Reports} {\bf 12} (1974) 75. See also
J. Polchinski, \npb 231, 1984, 269.

\refis{JJ1}I. Jack and D. R. T. Jones, \pl B234, 1990, 321; \npb 342, 1990,
127;
M. S. Al-sarhi, I. Jack, and D. R. T. Jones, \npb 345, 1990, 431.

\refis{JJ2}M. S. Al-sarhi, I. Jack, and D. R. T. Jones, ``Quadratic Divergences
in Gauge Theories'', preprint LTH266, July 1991.

\refis{Twoloop}G. Kleppe and R. P. Woodard, ``Two Loop Calculations Using
Nonlocal Regularization'', preprint VPI-IHEP-91/4, to appear in {\sl Ann.
Phys.}

\endreferences

\head{Figure Captions}

Figure 1: Contributions to the scalar two-point function.

Figure 2: Fermion two-point function graph.

Figure 3: Yukawa coupling renormalization graph.

Figure 4: Scalar quartic coupling renormaliztion graphs.

Figure 5: Contributions to the leading two loop quadratic divergence.

Figure 6: Two loop quadratic divergences from one loop renormalizations.

Figure 7: Two loop contribution due to a measure factor.

Figure 8: Quadratic divergence from Higgs self-interaction.

Figure 9: Quadratic divergence from hypercharge interaction.

Figure 10: Quadratic divergence from SU(2) interaction.

Figure 11: Quadratic divergence from top quark Yukawa coupling.

Figure 12: Quadratic divergence from hypercharge four-point interaction.

Figure 13: Quadratic divergence from SU(2) four-point interaction.

\end